# High Performance Field-Effect Transistor Based on Multilayer Tungsten Disulfide

Xue Liu[†¶], Jin Hu[†¶], Chunlei Yue[†], Della Fera, Nicholas D[†], Yun Ling[‡§], Zhiqiang Mao[†], Jiang Wei[†*]

[†] Department of Physics and Engineering Physics, Tulane University, USA

[‡] School of Materials Science and Engineering, Jiangsu University, China

[§] College of Electronic and Information Engineering, SuZhou University of Science and Technology, China

[¶] These authors contribute equally to this work
[*] Address correspondence to jwei1@tulane.edu

ABSTRACT:

Semiconducting two-dimensional transition metal chalcogenide crystals have been regarded as the promising candidate for the future generation of transistor in modern electronics. However, how to fabricate those crystals into practical devices with acceptable performance still remains as a challenge. Employing tungsten disulfide multilayer thin crystals, we demonstrate that using gold as the only contact metal and choosing appropriate thickness of the crystal, high performance transistor with on/off ratio of $10^8$ and mobility up to 234 $cm^2V^{-1}s^{-1}$ at room temperature can be realized in a simple device structure. Further low temperature study revealed that the high performance of our device is caused by the minimized Schottky barrier at the contact and the existence of a shallow impurity level around 80meV right below the conduction band edge. From the analysis on temperature dependence of field-effect mobility, we conclude that strongly suppressed phonon scattering and relatively low charge impurity density are the key factors leading to the high mobility of our tungsten disulfide devices.

Semiconducting transition-metal dichalcogenide (TMD) $MX_2$ with the thickness of atomic level, where M represents transition metal (Mo, W) and X represents chalcogen (Se, S), has attracted a lot of attention recently due to the new emerging electrical and optical properties and its great potential in practical applications. Its atomically thin structure can be easily obtained via techniques such as the micromechanical cleavage[1,2] and liquid-phase exfoliation,[3-5] because of the weak van der Waals bonding between neighboring chalcogen/metal/chalcogen layers. Not only bulk semiconducting TMD has the ideal size of band gap (1.1-2eV) for the making of transistor,[6] but also the band gap transforms from indirect to direct when the thickness of crystal approaches single atomic layer.[7-11] Consequently, significant enhancement of photoluminescence[8,9] has been observed in monolayer $MoS_2$.[7] Ultrasensitive photodetectors[12] based on $MoS_2$ has also been demonstrated. Moreover, the intrinsically broken valley degeneracy in monolayers makes TMD the perfect model system for studying "valleytronics".[13-15] Besides the exciting optical properties, the semiconducting TMDs show excellent electrical switching properties and, therefore, is expected to hold promise in modern nano-electronics applications. The electron mobility based on single atomic layer $MoS_2$ field-effect transistors (FET) ranges from ~1$cm^2V^{-1}s^{-1}$ [2] to ~200$cm^2V^{-1}s^{-1}$.[16] $MoS_2$ FET with $10^8$ on/off ratio has been achieved by using high-$\kappa$ oxide materials[16, 17] as the top gate dielectric grown by using atomic layer deposition (ALD). However, the challenges of obtaining monolayer, complicated device fabrication process and chemically modified interface make it difficult to study the intrinsic properties. Besides $MoS_2$, researchers are also interested in other members of this semiconducting TMD group such as $WSe_2$[18-20] and $MoSe_2$,[21, 22] which exhibit good switching behavior as well. Compared with $MoS_2$, $WSe2$ and $MoSe2$, tungsten disulfide ($WS_2$) has a larger band gap in both bulk crystal (~1.3eV) and monolayer (~2.1eV). Although $WS_2$ is expected to have the best transistor performance (the highest on-state current density and mobility) among all semiconducting TMDs according to earlier theoretical calculation,[23] only a few FET studies[24, 25] on $WS_2$ atomic crystal have been reported with the reached on/off ratio of $10^5$ and electron mobility of $20 cm^2 V^{-1} s^{-1}$, and yet, a device with performance comparable to that of $MoS_2$ has not been realized. In this paper, we report our studies on multilayer $WS_2$ field-effect transistor with simple back-gate device structure. Our devices exhibit exceptionally high on/off ratio reaching $10^8$ and mobility up to $234 cm^2 V^{-1} s^{-1}$. We have further performed low temperature measurements and revealed that minimized Schottky barrier at the contacts and appropriate impurity level plays critical roles in our $WS_2$ device.

**RESULTS AND DISCUSSIONS**

The example of as fabricated device is shows in Figure 1a and b. In this report, devices with various $WS_2$ crystal thicknesses ranging from 3nm to 100nm were studied. To characterize the electrical properties of the device, we use conventional three-terminal configuration setup with heavily doped silicon as the back-gate electrode, as illustrated by the schematic drawing in Figure 1c. Based on our test on more than one hundred devices, we find that $WS_2$ crystals with thickness around 6-10nm exhibit the best switching behavior (see supporting information for more details). Similar thickness dependence in $MoS_2$ has also been reported by Li et.al.[26] A typical measurement of source-drain current $I_{ds}$ vs. back gate voltage $V_{bg}$ at room temperature is shown in Figure 2a. The device appears to be *n*-type FET with the neutrality point of -45V. As shown in the log scale plot of $I_{ds}$-$V_{bg}$, the device can be tuned into an off-state with the minimum current $I_{ds}$ less than $10^{-14}$ A when $V_{bg}$ is set in the range of -30V to -50V. Also, the device have a weak ambipolar FET behavior with the sign of p-type carriers appearing at negative gate voltage ($V_{bg}$<-50V). The current $I_{ds}$ at on-state can reach 20μA at $V_{bg}$ ~ 60V with source-drain bias $V_{ds}$=1V. The on/off ratio at room temperature for this device reaches $10^8$, which is extremely high compared with the earlier results[17, 19, 24, 25] from other semiconducting members of the $MX_2$.[16,18,24,25] This is a surprising result given that we only use the device structure of back gating with 300nm $SiO_2$ as the dielectric layer. According to the gate sweep, we can extract the gate voltage induced carrier density $n_{2D}$ using the parallel-plate capacitor model $n_{2D} = C_{ox}(V_{bg} - V_{bg,th})/e$, where $C_{ox}$ is the dielectric capacitance per unit area and $V_{bg,th}$ is the threshold voltage and $e$ is the unit charge. $C_{ox}$ can be further calculated from $C_{ox} = \varepsilon_0 \varepsilon_r / d_{ox}$, where $\varepsilon_0$ is the dielectric constant of vacuum, and $\varepsilon_r$ represents the relative dielectric constant of 3.9 for $SiO_2$. The carrier density is estimated to be $2.6 \times 10^{12}\ cm^{-2}$ when $V_{bg}$=60V and $V_{ds}$=1V. Figure 2b shows the trace and retrace sweeps of $I_{ds}$-$V_{bg}$ with $V_{ds}$ fixed at 0.1V. The maximum voltage difference between trace and retrace sweeps is less than 1V indicating a small hysteresis in our device. This observation is in contrast to the relatively large hysteresis reported in other multilayer transition metal chalcogenides devices. Multiple factors, such as water or oxygen molecules absorbed on the surface[27-29] or charge injection at the oxide dielectric interface[30], should be considered as the possible causes. Figure 2c shows the $I_{ds}$-$V_{ds}$ characteristics at room temperature for different $V_{bg}$. At low source-drain bias from -0.1V to 0.1V, the $I_{ds}$-$V_{ds}$ presents near-linear and symmetric behavior indicating good contacts formed between gold contact metal and $WS_2$ crystal. However, when source-drain bias is swept up to 5V, as shown in Figure 2d, the $I_{ds}$ starts to display an upturn at $V_{ds}$=1V, and no signs of current saturation appears even when $V_{ds}$ is increased to 5V. To extract the room temperature field-effect mobility µ of our device, we utilize the field-effect transistor model $\mu = \left[\frac{dI_{ds}}{dV_{bg}}\right] \times [L/(WC_{ox}V_{ds})]$, where $L$ and $W$ are the channel length and width respectively. At room temperature, the extracted mobility of this particular devices corresponding to Figure 2 reaches $234 cm^2V^{-1}s^{-1}$ at $V_{bg}$–$V_{th}$ =25V (here $V_{th}$ is the threshold voltage), which is significantly higher than the mobility (typical value of 10 to 180$cm^2V^{-1}s^{-1}$) of MoS$_2$ devices with SiO$_2$ as the gate dielectrics.[2, 16, 17, 26, 31]

In order to reveal the origin of the high performance of our device, we have conducted low temperature transport measurement and studied the detailed transport mechanism on multiple devices. A device with channel dimensions of 8nm in thickness, 6µm in length and 3µm in width is selected for the following discussion. As shown in Figure 3a, the $I_{ds}$-$V_{ds}$ sweeps are recorded with back gate voltage fixed at $V_{bg}$=60V at different temperatures. It is apparent that the nonlinearity of $I_{ds}$-$V_{ds}$ is enhanced when the device has been cooled from 300K to 5K. In particular, unlike the relatively linear behavior of $I_{ds}$-$V_{ds}$ sweep at room temperature, the low temperature $I_{ds}$-$V_{ds}$ sweeps ($T$=200K to 5K) exhibit a distinct new feature: the device shows a relatively insulating state in the low source-drain bias regime (-0.6V<$V_{ds}$<0.6V) and a conducting state at large bias regime ($V_{ds}$>0.6V or $V_{ds}$ <-0.6V). The crossover source-drain voltage between the two regimes is around 0.6V. Above the crossover, $I_{ds}$ increases rapidly with $V_{ds}$ as can be seen more clearly in the insert of Figure 3a. We conjecture that the different temperature dependence behavior at small and large source-drain bias regimes may be caused by the existence of Schottky barrier, a typical situation for contacts between metal and semiconducting TMDs,[32] yet rarely explored in WS$_2$ device. Therefore, to highlight the possible Schottky barrier effect, we focus on the temperature dependence of source drain current at large $V_{bg}$ (60V to 35V) with small $V_{ds}$ (0.02V), as shown in Figure 3b. With large $V_{bg}$, the WS$_2$ crystal is tuned into the conducting on-state as illustrated by the band diagram of Figure 4a. Thus, the total conductance of the device will be limited only by the Schottky contacts. In Figure 3b, the $I_{ds}$ shows a near-exponential temperature dependence from room temperature to 90K, indicating a thermally activated charge transport mechanism. To accurately extract the Schottky barrier height of our device, we used the two dimensional (2D) thermionic emission transport equation $I_{ds} = A^* w T^{\frac{3}{2}} \exp\left(\frac{-q\phi_B}{k_B T}\right) \left\{\exp\left(\frac{qV_{ds}}{k_B T}\right) - 1\right\}$ to fit our data, where $A^*$ is the 2D Richardson constant, $w$ is the physical width of FET, $q$ is the unit charge, $\phi_B$ is the effective Schottky barrier height and $k_B$ is the Boltzmann constant.[33] By using the flat band voltage condition of $V_{bg}$=$V_{FB}$,[32, 34] the actual Schottky barrier height $\phi_{Sb}$ can be extracted to be 27.2meV for this particular device as shown in the insert in Figure 3b (see more extraction examples in Supporting Information). This barrier height is comparable to the thermal energy ($k_B T$~26meV) at room temperature. Therefore, the electrons at the Fermi level can be easily excited thermally to overcome Schottky barrier formed in our WS$_2$ device, which explains the near-linear $I_{ds}$-$V_{ds}$ curve at room temperature.

Intrinsic WS$_2$ crystal has an indirect band gap of 1.35eV[11, 35], which in general leads to a poor gatability using thick dielectric layer of 300nm SiO$_2$. However, the high on-off ratio in our multilayer WS$_2$ crystal indicates the opposite. This unusual behavior may suggest that our multilayer WS$_2$ crystal is fundamentally different to intrinsic WS$_2$. For conventional semiconductor, doping is the most common method to enhance the gatability of device[36, 37] by modulating the Fermi level within the band gap. Hence, we suspect our multilayer WS$_2$ crystal may also contains impurity. In order to set this impurity effect apart from the Schottky barrier effect, we measured the temperature dependence of conductance at a large source drain bias $V_{ds}$=2V with back gate voltage varied from -15V to 0V as shown in Figure 3c. The purpose of applying large source drain voltage up to 2V is to supply enough forward bias to lower the height of Schottky barrier at the drain side (illustrated by the band diagram in Figure 4b). Consequently, this favors the electron's transport toward drain electrode and minimizes the contact resistance caused by Schottky barrier. However, it is noteworthy that, because the $WS_2$ crystal is also tuned into the off-state by setting gate in the range of -15V to 0V, the charge carriers, i.e. electrons in this case, are mostly created through thermal excitations and we assume this is the major effective mechanism led to the conductance of our device in this configuration. Of course, if both gate voltage and source drain voltage are set to the configuration that $WS_2$ crystal is in on-state and Schottky barriers are minimized at the contacts, the device will be turned into a highly conductive state. This situation can be easily seen in the band diagram of Figure 4c. Now, we focus on the configuration that enables us to study the thermal activation behavior, i.e. large source drain bias and Vg=-15V to 0V. Figure 3c shows the temperature dependence of conductance, which indeed changes exponentially from 300K to 100K. We can fit the conductance data $G$ with the thermal activation equation of $G = G_o e^{-E_a/k_B T}$ that describes the temperature dependence of conductance for thermally activated charge carriers. Here $G_o$ is the conductance limit at high temperature and $E_a$ is the thermal activation energy. As shown in the insert of Figure 3c, the extracted thermal activation energy $E_a$ ranges from 80meV to 35meV, which is consistent with the value also been reported in $MoS_2$.[16] The obtained activation energy implies that the energy level generated by impurities is near the bottom of conduction band for our multilayer $WS_2$ crystal. Yet, the origin of impurity level is still under investigation and is one of the focuses in future work.

To further understand how the transport of charge carriers affected by the impurities or other possible scattering sources, we studied the temperature dependence of field effect mobility $\mu$. Figure 5a shows the on-state mobility at $V_g$-$V_{th}$=25V plotted as a function of temperature at different source-drain bias ($V_{ds}$=0.1V−2V). At small source-drain bias below 0.6V, mobility is underestimated due to the apparent Schottky barrier. At large source-drain bias ($V_{ds}$ →2V), contact resistance becomes negligible, which leads to more accurate extraction of mobility. On cooling from room temperature, the mobility follows a power law ($\mu \sim T^{-\gamma}$), and then remains at a saturated value below 100 K. The extracted power law constant $\gamma$ converges towards 1.15 at $V_{ds}$ =2V as shown in the insert of Figure 5a.

Now we focus on the detailed charge scattering mechanism of our device, which plays critical role in the FET performance of our devices. Similar to other TMD systems like $MoS_2$,[38] it is reasonable to consider that charge carriers in multilayer $WS_2$ crystal are usually scattered by acoustic phonon, optical phonon and charged impurities. For optical phonon scattering,[16, 39] the temperature dependence can be described with the equation $\mu_{op} = \frac{4\pi\varepsilon_0 \varepsilon_p \hbar^2}{e\omega m^{*2} Z_0}\left[e^{\frac{\hbar\omega}{k_B T}} - 1\right]$, where $\frac{1}{\varepsilon_p} = \frac{1}{\varepsilon_\infty} - \frac{1}{\varepsilon_s}$ with $\varepsilon_\infty$ and $\varepsilon_s$ are the high frequency and static dielectric constant respectively. $m^*$ is the effective mass of electron, $\hbar\omega$ is the optical phonon energy and $Z_0$ is the crystal thickness. For acoustic phonon scattering, under the deformation potential approximation, the mobility limited by acoustic phonons can be described by $\mu_{ac} = \frac{e\hbar^3 \rho v}{(m^*)^2 \Xi_\lambda^2 k_B T}$, where $\rho$ is the crystal density, $\Xi_\lambda$ is the deformation potential and $v$ is the acoustic phonon velocity.[38, 40] We shall include the background charge impurity scattering, which contributes to the mobility in the form[41]

$$\mu_{imp} = \frac{8\pi\hbar^3\varepsilon^2 k_F^2 \int_0^\pi \frac{\sin^2\theta}{(\sin\theta+\beta)^2}d\theta}{e^3 m^{*2} N_{imp}}.$$ Here $\beta = \frac{S_0}{2k_F}$ and $k_F$ is the wavevector on the Fermi surface, $N_{imp}$ is the impurity density and $S_0$ is the screening constant which takes the form of $S_0 = \frac{e^2 m^*}{2\pi\varepsilon\hbar^2}$.[42] Then the total mobility of a device was calculated using Mathiessen's rule $\mu^{-1} = \mu_{op}^{-1} + \mu_{ac}^{-1} + \mu_{imp}^{-1}$ by combine contributions from all three scattering sources. Using this model, we can fit the measured mobility (circles in Figure 5b) by changing variables of optical phonon energy $\hbar\omega$, deformation potential $\Xi_\lambda$ and charge impurity density $N_{imp}$ (other parameters were directly adopted from literatures of bulk $WS_2$[43, 44]). Figure 5b shows the fitted result (black curve), which agrees well with the data from our measurement. We extracted the optical phonon energy about $\omega \sim 240 cm^{-1}$, which is close to the theoretical value[43] of 300 cm$^{-1}$. The extracted deformation potential $\Xi_\lambda$ is around 2.0eV which is also close to the value of $MoS_2$.[38] The extracted charged impurity density of our $WS_2$ crystal is around $2 \times 10^9$ cm$^{-2}$, which is relatively small compared with other 2D crystals ($MoS_2$ with $1.8 \times 10^{10} cm^{-2}$ and grapheme with $\sim 10^{11} cm^{-2}$ ).[38, 45] From the fitted plot we conclude: 1. Charge impurity dominates the mobility at low temperature below 100K. 2. The acoustic phonon contribution to the mobility (blue curve) is marginal at all temperatures. 3. Optical phonon limits the mobility at temperatures above 100K. In particular, from 300K to 100K a power law temperature dependence of $\mu \sim T^{-\gamma}$ can be fitted with $\gamma$ approaching 1.15 at $V_{ds}$=2V as mentioned earlier. Comparing to the results from previously reported $MoS_2$ devices,[16] where $\gamma$=1.4 for back-gated device and 0.55 for dual-gated (both top-gate and back-gate are used) device, $\gamma$=1.15 from our $WS_2$ device indicates that optical phonon mode quenching may occur. It has been shown by earlier studies that top gate using high-$k$ dielectrics[2, 17, 46-48] is the reason to cause the suppression of optical phonons and screening of charged impurities and to improve the device mobility on $MoS_2$. However, in our $WS_2$ device the mobility is surprisingly high (up to 234 cm$^2$V$^{-1}$s$^{-1}$) without such top gate/-$k$ dielectric structure. This may imply that simple structure like a few layer $WS_2$ crystals on $SiO_2$ surface can also have optical phonon suppressed by other reasons. The overall charge scattering behavior of our $WS_2$ device appears similarly to that of two dimensional electron gas (2DEG) system containing low density and high mobility of electrons, such as AlGaN/GaN field-effect transistor.[49]

**CONCLUSIONS**

In conclusion, we have successfully fabricated back gated field-effect transistors based on multilayer $WS_2$ crystals with thermal $SiO_2$ as the dielectric layer and pure gold as the contact metal. These devices exhibit on/off ratios up to $10^8$ and mobilities reaching 234cm$^2$V$^{-1}$s$^{-1}$ at room temperature. We found that the device performance strongly depends on the thickness of $WS_2$ channel crystal and the best performance appears in the thickness between 6 to 10nm. By studying low temperature transport behavior, we have identified that the high performance of our device is a combined result of low contact Schottky barrier and shallow impurity level inside $WS_2$ crystal. We also revealed that the scattering of charge carriers is mainly caused by optical phonons and charge impurities. Our work demonstrates that, comparing to other semiconducting TMDs, multilayer $WS_2$ crystal is indeed an excellent channel material for the making of high performance FET required for both energy saving and high power electronics applications.

**METHODS**

The high quality $WS_2$ bulk single crystals were synthesized using chemical vapor transport[50] with iodine acting as transport agent. The standard micromechanical exfoliation technique[2] was adopted to obtain

WS$_2$ thin flakes on the substrate of silicon with 300nm thermal oxide. The flakes with desired thicknesses were screened by using an optical microscope via color contrast. Then 35nm gold metal contacts connecting the WS$_2$ flakes were patterned and deposited by using standard electron beam lithography and thermal evaporation. Then, the devices were annealed in forming gas at 200$^o$C for 2 hours in order to remove organic residues introduced during the fabrication process. The accurate thickness of WS$_2$ flakes were further determined by using Atomic Force Microscope (AFM).

*Acknowledgment.* We acknowledge Cleanroom Facility and Coordinated Instrumentation Facility at Tulane University for technical support of scanning electron microscopy and micro/nanofabrication.

*Supporting Information Available:* Mobility histogram; Schottky barrier height extraction. This material is available free of charge via the Internet at http://pubs.acs.org.

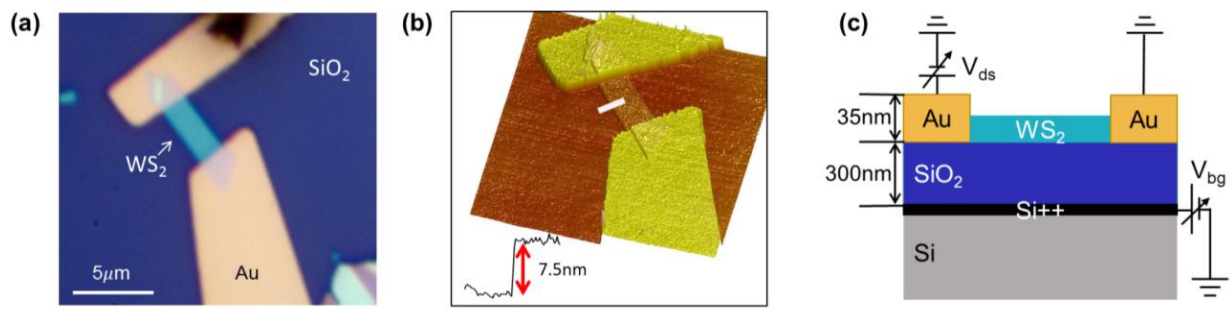

**Figure 1.** Device fabrication of $WS_2$ FET. (a) Optical image of a multilayer $WS_2$ device with gold as the contact metal and $Si/SiO_2$ as the substrate. (b) Atomic force microscope image of the device shown in (a). The insert shows the line trace of 7.5 nm thick $WS_2$ crystal obtained at the location of white line. (c) Schematic drawing of the measurement setup for electrical transport characterization.

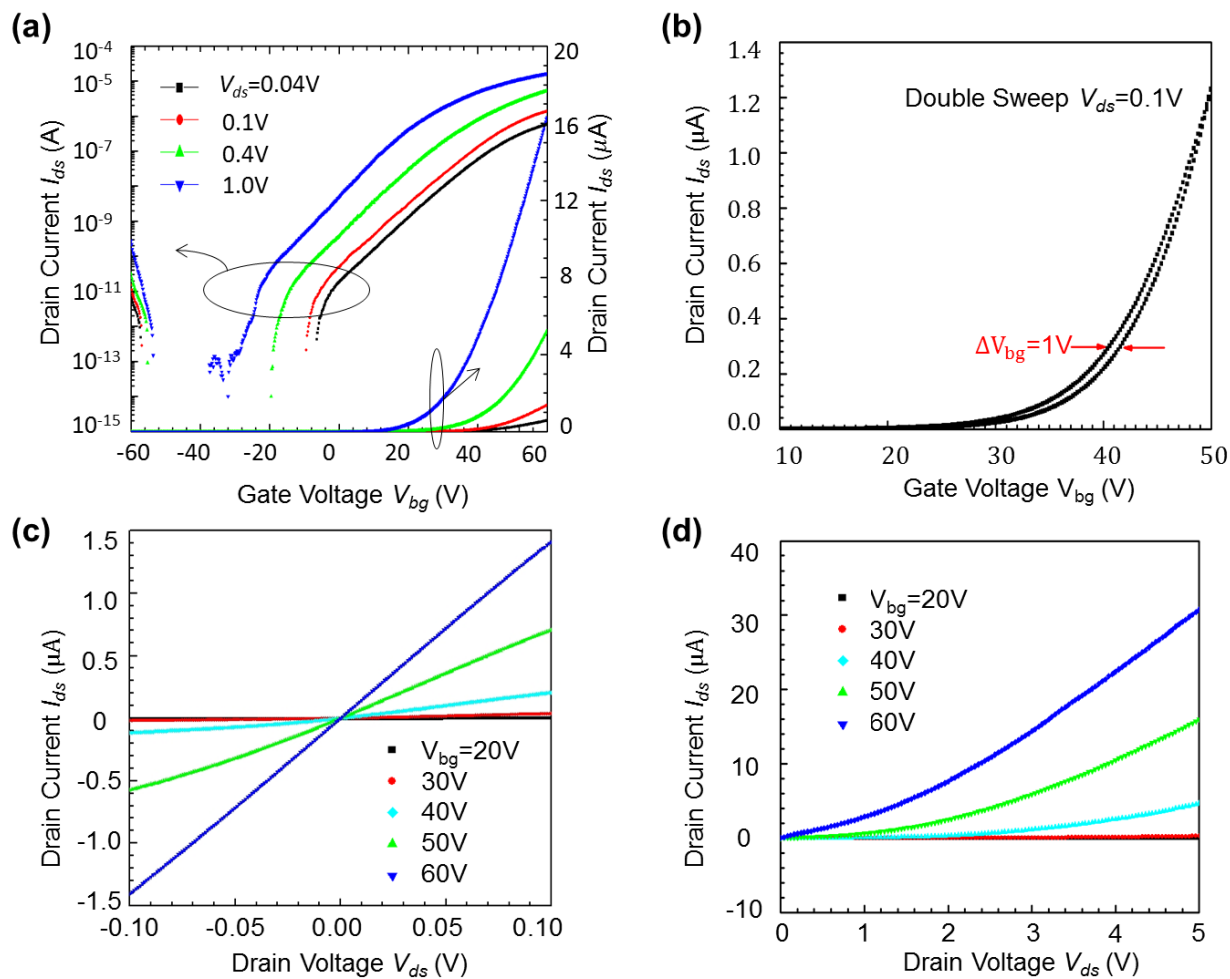

**Figure 2.** Electrical property characterization of $WS_2$ device (shown in Figure 1) at room temperature. The lateral dimension is 4.8um in length and 1.8um in width. (a) $I_{ds}$-$V_{bg}$ sweeps under different source drain voltages $V_{ds}$=0.04V, 0.1V, 0.4V, 1V plotted in log scale (left vertical axis) and linear scale (right vertical axis). Note: for this particular device, the maximum $I_{ds}$ on/off ratio reaches $10^8$ at $V_{ds}$=1V and the mobility extracted reaches $234 cm^2 V^{-1} s^{-1}$ at room temperature. (b) Trace and retrace $I_{ds}$-$V_{bg}$ sweep at $V_{ds}$=0.1V showing small hysteresis at room temperature. (c) $I_{ds}$-$V_{ds}$ sweeps in samll source-drain bias regime (-0.1 to 0.1V) at room temperature, showing near-linear behavior. (d) $I_{ds}$-$V_{ds}$ sweeps in large source-drain bias regime (0 to 5V), exhibiting clear nonlinearity.

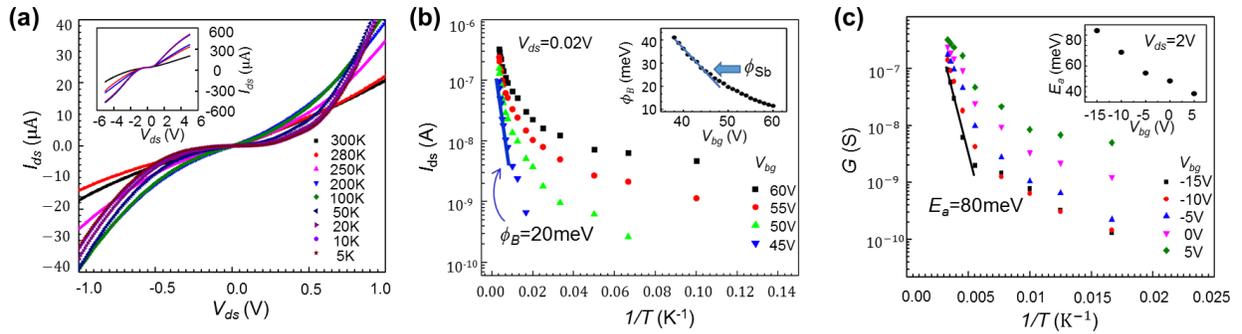

**Figure 3.** Low temperature characterization. (a) $I_{ds}$-$V_{ds}$ sweeps in the range of $V_{ds} = \pm 1V$ at on-state ($V_{bg}$=60V) under different temperatures from 300K to 5K. Insert: zoomed out $I_{ds}$-$V_{ds}$ sweeps in the range of $V_{ds} = \pm 6V$. (b) Source drain current $I_{ds}$ (in log scale) vs. temperature T (in reciprocal scale) for the same device in (a) with $V_{ds}$=0.02V and $V_{bg}$=45V~60V. Insert: the extracted effective Schottky barrier height $\phi_B$ and the arrow indicates the actual Schottky barrier height $\phi_{sb}$. (c) Conductance G (in log scale) vs. temperature T (in reciprocal scale) for the same device in (a) with $V_{ds}$=2V and $V_{bg}$=-15~5V. Insert: the extracted thermal activation energy $E_a$.

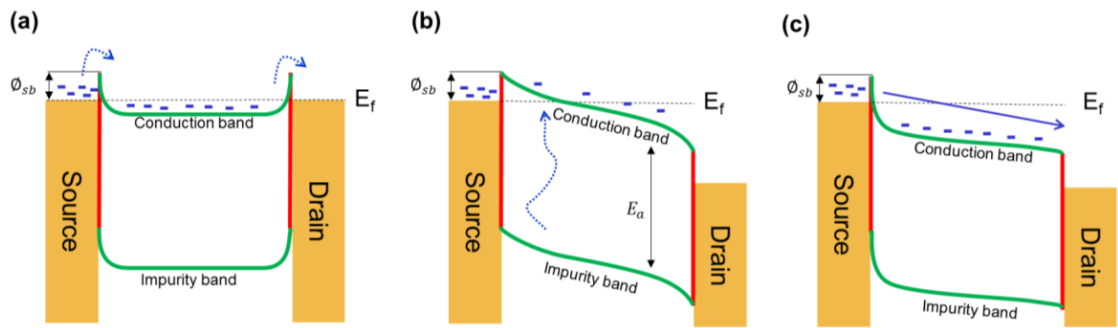

**Figure 4**. (a) Band diagram of a $WS_2$ device operating at low temperature and low source-drain bias, when the $WS_2$ crystal is set to on-state with appropriate gate voltage. This diagram illustrates that at low temperature, Schottky barrier becomes the major factor to limit the source-drain current. (b) Band diagram of a $WS_2$ device operating at low temperature and large source-drain bias, when the $WS_2$ crystal is set to off-state with appropriate gate voltage. This diagram illustrates that at low temperature thermal excitation between impurity band and conduction band are the main mechanism generating charge carriers. (c) Band diagram of a $WS_2$ device operating at room temperature with large source-drain bias, while the $WS_2$ is set to on-state with appropriate gate voltage. (Note: all the band diagrams are only for the purpose of illustrations and not drawn to scale.)

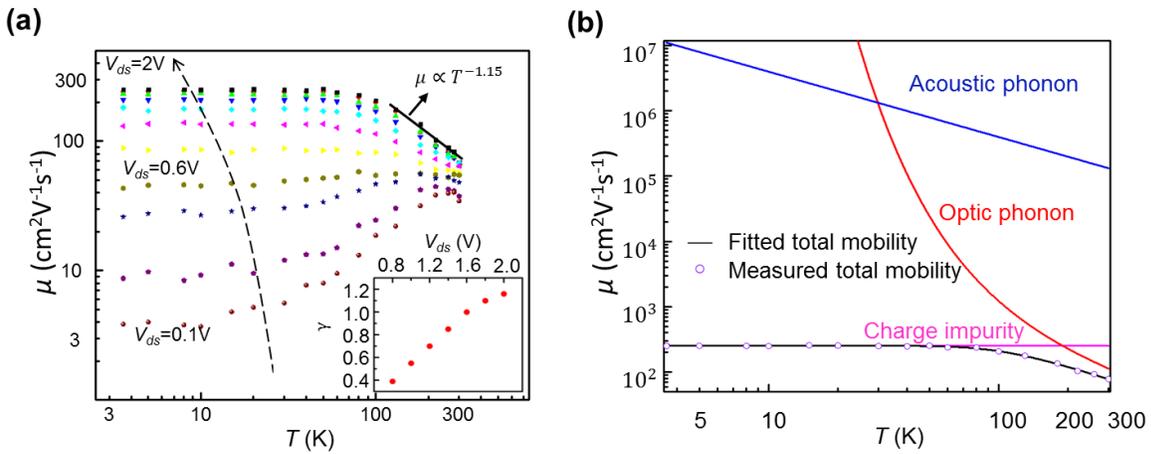

**Figure5**. (a) Temperature dependence of field-effect mobility for a WS$_2$ device at different source-drain bias ( 0.1V to 2V) with fixed effective gate voltage($V_{bg}$-$V_{th}$=25V). Insert: the power law constant $\gamma$ vs. source-drain bias (extracted from (a)). Dimension of the device: thickness 7.5nm, channel length 6μm and channel width 3μm. (b) Temperature dependence of measured electron mobility (hollow circular dots) at $V_{ds}$=2V and the fitted electron mobility (black line). The fitted data includes the scattering contribution from acoustic phonons (blue line), optical phonons (red line) and charge impurities (magenta line).